\documentclass[structabstract]{aa}
\usepackage{graphicx}
\usepackage{url}
\usepackage{natbib}
\usepackage{txfonts}

\newcommand{\dr}{UKIDSS-GPS DR3}
\newcommand{\nir}{NIR}
\newcommand{\xmm}{\emph{XMM-Newton}}
\newcommand{\puls}{2XMM~J174016.0$-$290337}
\newcommand{\sext}{{\tt SExtractor}}
\newcommand{\jh}{\ensuremath{J\!-\!H}}
\newcommand{\jk}{\ensuremath{J\!-\!K}}
\newcommand{\hk}{\ensuremath{H\!-\!K}}

\begin{document}

   \title{A New 626 s Periodic X-ray Source in the Direction of the Galactic Center}

   \author{S. A. Farrell
    \inst{1, 2}
          \and
          A. J. Gosling
           \inst{3, 4}
           \and
          N. A. Webb
           \inst{1}
          \and
          D. Barret
           \inst{1}
            \and
          S. R. Rosen
           \inst{2}
            \and
          M. Sakano
           \inst{2}
                       \and
          B. Pancrazi
           \inst{1}
          }

   \institute{Universit\'{e} de Toulouse, UPS, CESR, 9 Avenue du Colonel Roche, F-31028 Toulouse Cedex 9, France\\
\and Department of Physics and Astronomy, University of Leicester, University Road, Leicester, LE1 7RH, UK\\
\email{saf28@star.le.ac.uk} 
\and
Astronomy Division, Department
of Physical Sciences, PO Box 3000, FIN-90014 University of Oulu,
Finland\\
\and Department of Astrophysics, University of Oxford, Keble Road, Oxford, OX1 3RH, UK}

   \date{Received ****; accepted ****}


  \abstract
   {}
   {Here we report the detection of a 626 s periodic modulation from the X-ray source \puls\ located in the direction of the Galactic center.}
   {We present temporal and spectral analyses of archival
     \emph{XMM-Newton} data and photometry of archived near-infrared
     data in order to investigate the nature of this source.}
   {We find that the X-ray light curve shows a strong modulation at 626 $\pm$ 2 s with a confidence level $>$ 99.9$\%$ and a pulsed fraction of 54$\%$.  Spectral fitting
     demonstrates that the spectrum is consistent with an absorbed power law. No significant spectral variability was observed over the 626 s period. We have investigated the possibility that the 626 s period is orbital in nature (either that of an ultra-compact X-ray binary or an AM CVn) or related to the spin of a compact object (either an accretion powered pulsar or an intermediate polar).}
   {The X-ray properties of the source and the photometry of the candidate near-infrared counterparts are consistent with an accreting neutron star X-ray binary on the near-side of the Galactic bulge, where the 626 s period is most likely indicative of the pulsar spin period. However, we cannot rule out an ultra-compact X-ray binary or an intermediate polar with the data at hand. In the former case, if the 626 s modulation is the orbital period of an X-ray binary, it would be the shortest period system known. In the latter case, the modulation would be the spin period of a magnetic white dwarf. However, we find no evidence for absorption dips over the 626 s period, a low temperature black body spectral component, or Fe K$\alpha$ emission lines. These features are commonly observed in intermediate polars, making \puls\ a rather unusual member of this class if confirmed. Based on the slow period and the photometry of the near-infrared counterparts, we instead suggest that \puls\ could be a new addition to the emerging class of symbiotic X-ray binaries.} 
   
   \keywords{Accretion, accretion disks -- (Stars:) binaries: general -- (Stars:) pulsars: individual: \puls\ -- X-rays: binaries
               }

   \maketitle
%

\section{Introduction}

X-ray sources displaying periodic variability on timescales of $\sim$1000 s are typically associated with binary star systems containing an accreting compact object, where the modulation is indicative of either the spin of the compact object (in the case of a white dwarf or neutron star primary), or the orbit of an ultra-compact binary system. These systems are generally classified by the nature of the compact object: cataclysmic variables (CVs) are systems containing a white dwarf accreting from a low mass companion star, while X-ray binaries are systems containing either a neutron star or black hole accreting from either a low mass or high mass companion (LMXBs and HMXBs respectively). 

The majority of CVs have orbital periods between $\sim$1 --  8 h. A rare sub-class of CVs known as the AM CVn systems have orbital periods between $\sim$10 -- 60 min, where the orbit is so compact that only a white dwarf mass donor is allowed. This scenario is reached after the system passes through two common envelope phases, leaving both stellar cores exposed. CVs are split into two main classes: the non-magnetic systems with white dwarf magnetic field strengths $\la$ 10$^4$ G, and the magnetic systems with field strengths up to a few tens of MG. The magnetic CVs are further split into two sub-classes: intermediate polars (IPs) and polars. The  accretion discs in IPs are truncated by interactions with the magnetosphere of the white dwarf. The interactions between the white dwarf magnetic field and the accreting material in polar systems are much stronger, so that the material flowing from the donor couples directly onto the magnetic field lines of the white dwarf and no disc forms. The magnetic field also couples to the field of the donor star, forcing the white dwarf to co-rotate with the binary orbit. The rotation period of a white dwarf can manifest itself as an X-ray modulation only when its magnetic field strength is sufficiently high enough to channel the accreting material along the field lines onto the magnetic poles. If the magnetic and rotation axes are misaligned (and if the viewing angle allows), the X-ray emission arising from the polar regions can be modulated over the spin period (in a similar fashion to the ``light house" affect seen in neutron star pulsars). In polar systems the spin period and orbital period cannot generally be distinguished due to the co-rotation. IPs on the other hand do not co-rotate, so the spin and orbital periods can be observed as separate modulations. The spin periods in IPs typically fall in the range between $\sim$0.5 -- 142 min. See \citet{ku06} for a comprehensive review of CVs.

X-ray binaries with orbital periods less than $\sim$1 hr are a subset of LMXBs known as ultra-compact X-ray binaries (UCXBs). The short orbital period implies a very small Roche lobe and requires a hydrogen-poor donor star \citep{int07}. Globular clusters are host to five times more UCXBs than the field, likely due to the higher probability of stellar encounters \citep{int07}. The shortest ultra-compact orbital period belongs to the globular cluster source 4U 1820-303 with an orbital period of 685 s \citep{whi86}.

Accreting neutron star X-ray pulsars with spin periods in the range
of $\sim$200 -- 1500 s are typically identified with HMXBs, where the mass donor is either a Be or supergiant star \citep{liu06}. In these systems accretion takes place either via the passage of the neutron star through a dense circumstellar disc or via the stellar wind, and the slow spin periods can be explained due to a lower accretion efficiency providing less spin-up torque than generated by Roche lobe overflow. Typical LMXB pulsars have spin periods less than 200 s, with the majority of periods below 1 s \citep{liu01,rit09}. 

Recently, a new class of LMXB was discovered
with each source associated with an M-type giant. These \emph{symbiotic} X-ray binaries are believed to consist
of neutron stars in a wide orbit with the donor star, accreting via
the stellar wind of the M giant  \citep[e.g.][]{mas06a}. Only a handful of symbiotic X-ray
binaries have so far been discovered -- GX 1+4 \citep{dav77,cha97}, 4U 1700+24 \citep{gar83,mas02}, 4U
1954+319 \citep{mas06a,mas06b,mat06}, Sct X-1 \citep{kap07}, IGR J16194-2810 \citep{mas07}, and 1RXS J180431.1-273932 \citep{nuc07}. Of these systems only 3 are confirmed
to contain X-ray pulsars -- GX 1+4 \citep[P$_{spin}$ = 120 s;][]{lew71}, Sct
X-1 \citep[P$_{spin}$ = 112 s;][]{koy91}, and 1RXS J180431.1-273932 \citep[P$_{spin}$ =
494 s;][]{nuc07} -- though a $\sim$5 hr periodic modulation detected in the emission from 4U 1954+31 has also been suggested as possibly indicating the spin of the neutron star \citep{cor08}. 

In this paper we present the results of
analyses of archival \emph{XMM-Newton} EPIC data of the X-ray
source \puls , a variable source which exhibits a strong periodic modulation over 626 s. Following initial submission of this paper, a report of a hard X-ray counterpart to \puls\ detected by \emph{INTEGRAL} has appeared in the literature \citep{mal10}, including cursory analyses of the \xmm\ EPIC spectra presented in this paper. Other reports of the detection of a 623 $\pm$ 2 s period in the same \emph{XMM-Newton} data presented here \citep{hal10a} and an $R$-band optical photometric period of 622 $\pm$ 7 s \citep{hal10b} have also appeared, although little detail was given in these telegrams. Here we present a more detailed analysis of the \xmm\ data and a comprehensive discussion of the possible nature of \puls . In $\S$ 2 we introduce the 2XMM catalogue and briefly outline the method by which \puls\ was identified. In $\S$ 3 we describe the observations and data reduction steps. $\S$ 4 details the spectral and timing analyses performed on the data. A search of archival all-sky optical and near-infrared (NIR) data for candidate counterparts is outlined in $\S$5. $\S$ 6 provides
a discussion of our results, and $\S$ 7 presents the conclusions that
we have drawn.

\section{The 2XMM Catalogue}

The Second \emph{XMM-Newton} Serendipitous Source Catalogue \citep[2XMM;][]{wat08} was
released on 2007 August 22 and is the largest X-ray source catalogue
ever produced, containing almost twice the number of discrete sources
as the \emph{ROSAT} all-sky surveys \citep[RASS;][]{vog99,vog00}.  The 2XMM catalogue
contains 246,897 source detections drawn from 3,491 \emph{XMM-Newton}
European Photon Imaging Camera (EPIC) observations made between 2000
February 3 and 2007 May 01, representing 191,870 discrete sources. A
simple $\chi^2$ variability test against a null hypothesis of
constancy was applied to the flare-filtered binned light curves (typical binning
timescale $\gg$ 10 s) extracted by the pipe-line for each
source \citep[see][for a detailed discussion of the variability test]{wat08}. All sources which had $\chi^2$ $<$ 10$^{-5}$ from the
variability test applied to any of the EPIC light curves in any of the
observations processed were flagged in the catalogue as variable. A
total of 2,001 discrete variable sources were flagged in this manner. We have manually inspected the pipe-line produced images, spectra and
light curves for all 2,001 discrete variable sources in 2XMM (Farrell
et al. in preparation). During this investigation we identified \puls, a 2XMM source which exhibits significant periodic variability in the EPIC light curves.

\section{X-ray Observations}

\subsection{Data Reduction}
\subsubsection{\emph{XMM-Newton}}

The field of \puls\ was observed with $\xmm$ (ObsId: 0304220101) on
September 29th 2005 (MJD 53642) for 7.7 ks \citep[see][for a description of the \xmm\ spacecraft and scientific payload]{jan01}. The observation was performed with the three EPIC cameras
(pn, MOS1 and MOS2) in imaging mode with a medium filter.  The pn
camera was in small window mode while both the MOS cameras were in
full frame mode. The Optical Monitor (OM) observed the field in imaging mode (standard  RUDI-5 mosaic) with the \emph{uvw2} filter, while the Reflection Grating Spectrometers (RGS) observed in the standard spectroscopy mode. 

Source detection was performed by the \xmm\ pipeline \citep[see][for a description of the process]{wat08}.
\puls\ was detected at an off-axis angle of $\sim$1.3\arcmin\ at
right ascension 17h 40\arcmin 16.0\arcsec, declination -29$\degr$ 03\arcmin 38\arcsec\
(J2000), with a 3$\sigma$ positional uncertainty of 3\arcsec. This
uncertainty was derived by adding in quadrature the statistical error
obtained through fitting the point spread function (PSF) and the
systematic pointing error of 1\arcsec. Astrometric correction using optical or NIR catalogues was not applied, as the extremely crowded nature of the field made it impossible to reliably match X-ray and optical/NIR source positions. As the source does not fall on the pn CCD due to the off-axis
position and the pn small window mode, only MOS data are available for
this source. No source was detected in the OM mosaiced image within the EPIC error circle, and not enough counts were detected from \puls\ to produce a viable RGS spectrum. We therefore concentrate on the MOS data for our analyses.

The Observation Data Files (ODF) were processed using the $\xmm$ Science Analysis System v9.1
(SAS) software\footnote{http://xmm.esac.esa.int/sas/}. The MOS data were reduced using the $\tt{emproc}$ script with the most recent calibration data files. Bad
events due to bad rows, edge effects, and cosmic rays were flagged and
discarded. The resulting cleaned event lists were filtered for event
patterns in order to maximise the signal-to-noise ratio against non
X-ray events, with only calibrated patterns (i.e. simple to quadruple
events) selected.  Single event light curves with energies exceeding
10 keV were produced for each MOS camera in order to identify periods of high background
related to soft proton flares. A small flare was present in the light curves $\sim$1 ks into the observation, with a peak count rate of 0.8 count s$^{-1}$ in each camera. Good time interval (GTI) filtering was therefore applied to remove intervals with background rates above 0.2 count s$^{-1}$ for the extraction of spectra and light curves so as to maximise the signal-to-noise.

\begin{figure*}
   \begin{center}
   \includegraphics[width=0.9\textwidth]{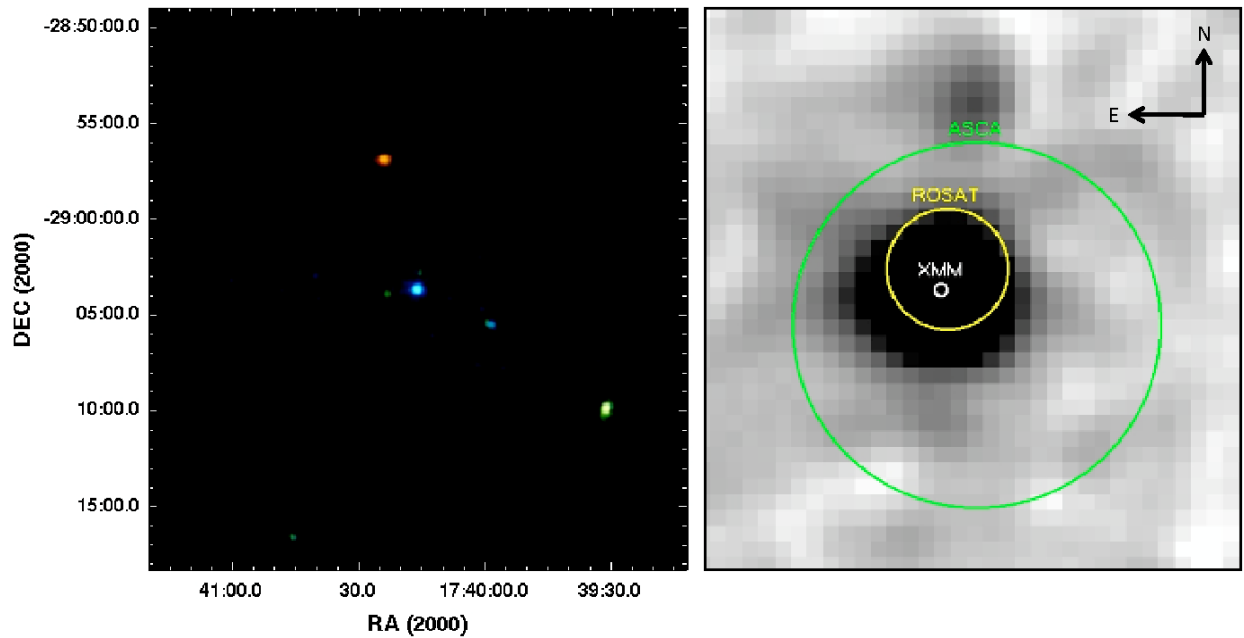}
  \caption{\emph{Left:} Combined (MOS1 $\&$ MOS2) EPIC RGB (Red: 0.2 -- 1 keV, Green: 1 -- 2 keV, Blue: 2 -- 10 keV) image showing a $\sim$28\arcmin\ $\times$ 28\arcmin\ field centered on \puls.
   The image has been Gaussian smoothed with a kernel radius of 3 pixels. \emph{Right:} EPIC image showing the positions of the previously detected X-ray sources AX J1740.2$-$2903 (ASCA, green circle), AX J1740.3$-$2904 (ROSAT, yellow circle) and the new \xmm~position (XMM, white circle). The circle radii represent the 90$\%$ errors in each case.}
              \label{epicim}
    \end{center}
    \end{figure*}

Images in three energy bands (Red = 0.2 -- 1 keV, Green = 1 -- 2 keV, and Blue = 2 -- 10
keV) were produced from the MOS1 and MOS2 data, and combined into a
merged RGB mosaic in order to derive spectral colours (Figure
\ref{epicim}). \puls\ is the brightest blue (i.e. hard) source in the
centre of the field. A number of fainter sources are also present in the
image, with the three brightest sources after \puls\ in the image (2XMM J174023.8-285652, 2XMM J173958.5-290529 and 2XMM J173931.2-290953 from left to right respectively in Figure \ref{epicim}) all coincident with moderately bright optical counterparts and with X-ray spectra consistent with thermal plasma models. They are therefore likely to represent emission from stellar coronae.

\subsubsection{\emph{ASCA} \& \emph{ROSAT}}

\puls\ lies within the 50\arcsec\ positional error circle (90\% confidence level) of AX J1740.2$-$2903, a hard X-ray source detected by $\it{ASCA}$ \citep{sak02}. The derived position is also 5.99\arcsec\ away from a point-like X-ray source in the catalogue of
soft X-ray sources in the Galactic center region compiled using
archived \emph{ROSAT} Position Sensitive Proportional Counter (PSPC)
data by \citet{sid01}, who designated it AX J1740.3$-$2904. The positional error of the PSPC detection is stated as 10\arcsec, although the confidence level is not given in \citet{sid01}. We can therefore only presume it is the same as used for the RASS \citep[i.e. 1$\sigma$;][]{vog99}. Examination of the combined MOS1 and MOS2 0.2--12 keV EPIC image finds no other source within the 90\% confidence error circles of either AX J1740.3$-$2904 or AX J1740.2$-$2903 (Figure \ref{epicim}). We therefore  conclude that \puls , AX J1740.3$-$2904 and AX J1740.2$-$2903 are all likely to be the same source.

The field of \puls~was observed with \emph{ASCA} on 19 September 1997 (ObsId: 44) and 7--8 September 1998 (ObsID: 58) with the GIS for exposures of 17 and 10 ksec respectively \citep{sak02}. The source was detected well off-axis in both observations, where the effective area is considerably smaller due to the vignetting effect. The data were reduced and filtered using the same method and criteria described in \citet{sak02}. 

The Galactic center region was observed numerous times by \emph{ROSAT} in a series of short raster scans. \puls~ was detected during one of these scans with an effective exposure time of $\sim$2 ks. The total PSPC 0.1 -- 2.4 keV count rate was 0.019 $\pm$ 0.003 count s$^{-1}$ \citep{sid01}, giving a total of only $\sim$40 photons during the observation, insufficient for meaningful timing or spectral analyses. We therefore concentrate on the \xmm\ and \emph{ASCA} data in the following studies.

\subsection{Source Extraction}

Events within a circular region of radius 80\arcsec\ around the position of
\puls\ were extracted from the EPIC data, with circular regions around two nearby faint X-ray
sources excluded (with radii of 35\arcsec\ and 22\arcsec\ for the brightest and faintest sources respectively) . Background events were extracted from a region
containing no other X-ray sources at the same off-axis position and
with the same area. Source and background light curves with the
maximum time resolution of 2.6 s were extracted for each MOS camera
and then combined into a single EPIC light curve to improve statistics
and enhance signal-to-noise. Background subtraction was performed and
the resulting combined EPIC light curve was then corrected to the barycentre using the task $\tt{barycen}$. Soft (0.2 -- 2 keV) and hard (2 -- 10 keV) band light
curves were extracted and combined using the same technique, so that a
comparison between the source variability in different energy bands
could be made. Source and background spectra were then extracted for
each camera and response and ancillary response files generated. The
spectra were grouped at 20 counts per bin to provide sufficient
statistics for spectral analyses.


\section{Timing \& Spectral Analyses}

\begin{figure}[h!]
   \begin{center}
   \includegraphics[width=\columnwidth]{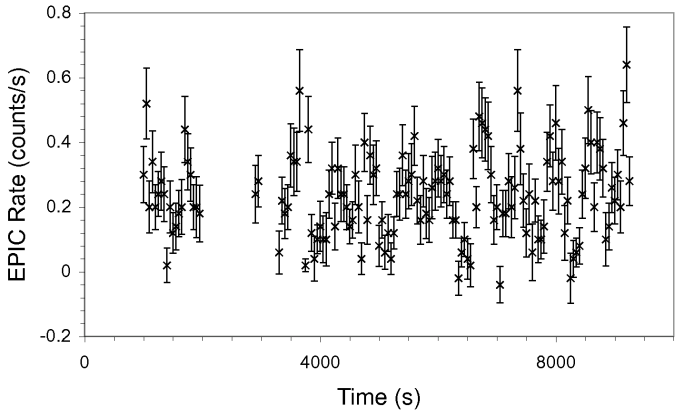}
   \caption{Combined (MOS1 $\&$ MOS2) 0.2 -- 10.0 keV EPIC background subtracted light curve
     binned at 50 s. For clarity, the time axis indicates the time
     since MJD 53642.034. The gaps in the light curve are a result of filtering out background flares.}
              \label{epiclc}
    \end{center}
    \end{figure}

\begin{figure}[h!]
   \begin{center}
   \includegraphics[width=\columnwidth]{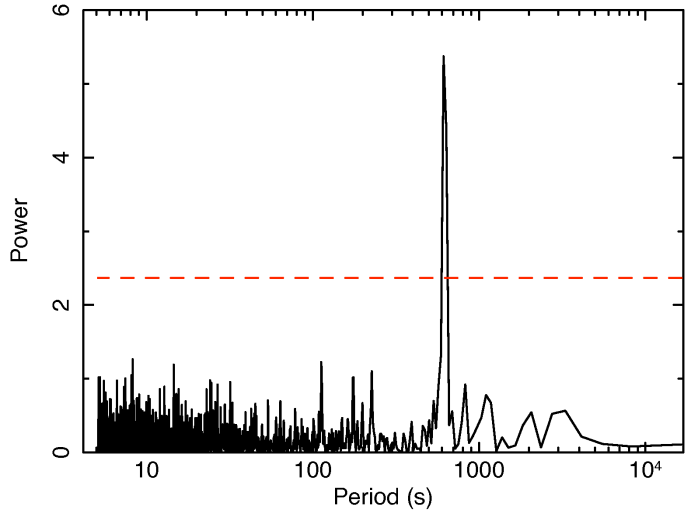}
   \caption{Lomb-Scargle power spectrum of the combined EPIC background subtracted light
     curve with background flares filtered out binned at 2.6 s showing the significant detection of a
     modulation at 626 s. The dashed line indicates the
     99.9$\%$ white noise significance level.}
              \label{powspec}
    \end{center}
    \end{figure}

Clear variability can be seen in the EPIC light curve shown in Figure
\ref{epiclc}, which was binned at 50 s for clarity. A power spectrum
was generated for the 2.6 s resolution EPIC light curve (Figure
\ref{powspec}) using the fast periodogram $\tt{fasper}$ subroutine of
the Lomb-Scargle periodogram \citep{lo75,sc82,pr89}, with the 99.9$\%$
white noise significance levels estimated using 1000 Monte Carlo
simulations \citep*[e.g.][]{ko98}. A highly significant peak at a
period of $\sim$630 s can be seen, indicating the presence of
significant periodic variability.  No other significant
peak down to the Nyquist period of 5.2 s was
detected. 

\begin{figure}[h!]
   \begin{center}
   \includegraphics[width=\columnwidth]{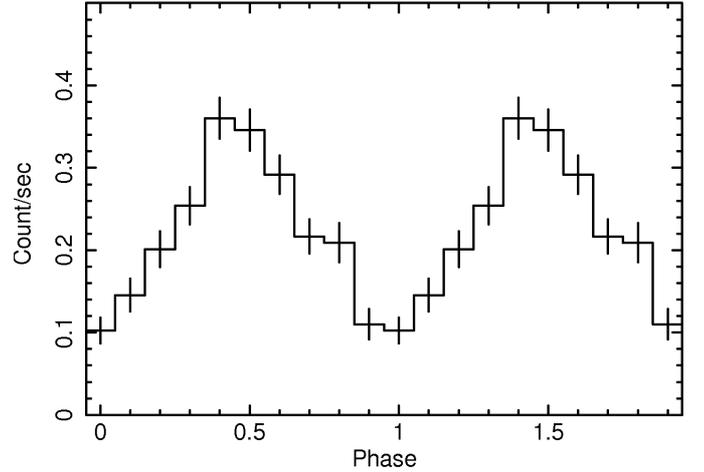}
   \caption{The EPIC 0.2 -- 10 keV background subtracted light
     curve folded over the best-fit period, with phase zero set at
     the minimum of the modulation. The error bars represent the 1$\sigma$ confidence level.}
              \label{tfold}
    \end{center}
    \end{figure}

In order to accurately constrain the
value of the period, we employed the epoch folding search technique. We performed a search using the HEASOFT task {\tt efsearch} for periods between 320 -- 920 s with a resolution of 1.2 s. The resulting period vs $\chi^2$ plot found a significant peak that when fitted with a Gaussian profile was found to be centered at 626 s. The EPIC 0.2 -- 10 keV light curve was then phase
folded over the best-fit period in order to examine the profile of the
modulation (Figure \ref{tfold}). The profile is consistent with a smooth sinusoidal function, therefore we used the method described by Equation 4 in \citet{lar96} for determining the period error for purely sinusoidal modulation. In this manner we constrained the period to 626 $\pm$ 2 s. To test the possibility that the true period is twice this value (possibly due to periodically viewing both poles of a magnetic white dwarf or neutron star), we folded the light curve over 1252 s. The shape of the profile between phases 0.0 -- 0.5 was entirely consistent with that between phases 0.5 -- 1.0 within the errors, compatible with a true period of 626 s. A search for periodic variability was also performed using the combined \emph{ASCA} GIS data in the 1--10 keV energy band, finding no evidence for significant periodic variability. This non-detection is likely due to poor statistics, as the total number of counts after background subtraction was only $\sim$560, with the estimated background level 50$\%$ higher than the source counts.

\begin{figure}[h!]
   \begin{center}
   \includegraphics[width=\columnwidth]{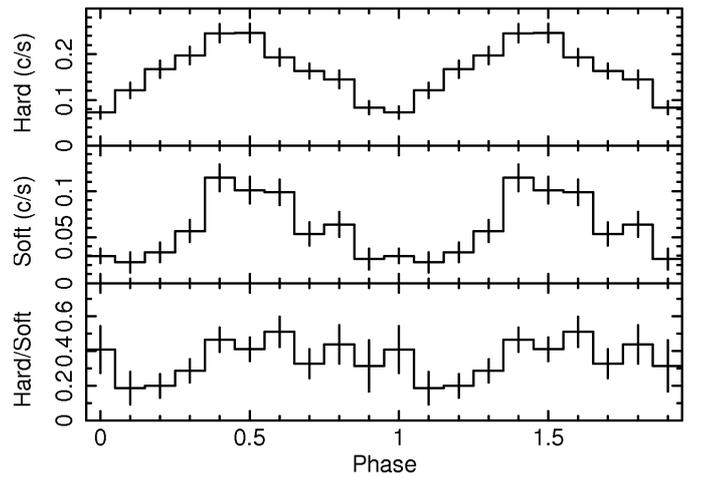}
   \caption{The EPIC 2 -- 10 keV (top) and 0.2 -- 2 keV (middle) background subtracted light
     curves folded over the best-fit period, with phase zero set at
     the minimum of the modulation in the total band folded light curve shown in Figure \ref{tfold}. The (0.2 -- 2 keV)/(2 -- 10 keV) hardness ratio is shown in
     the bottom panel. The error bars represent the 1$\sigma$ confidence level.}
              \label{folds}
    \end{center}
    \end{figure}

The EPIC hard and soft band light curves were also phase folded over the best-fit period in order to examine the variability of the profile with energy, with the (0.2 -- 2 keV)/(2 -- 10 keV) hardness ratio
calculated for each phase bin in order to crudely investigate spectral variability over the 626 s period (Figure \ref{folds}). The hardness ratio appears to vary slightly over the 626 s period, although none of the points deviate significantly from the average value within the errors. We therefore cannot draw any conclusions regarding spectral variability over the 626 s period using the hardness ratios. We investigate the possibility of spectral variability more quantitatively below. The pulsed fraction of the total band modulation,  defined as the ratio of the pulsed counts to the pulsed + non-pulsed counts \citep[e.g.][]{kui02}, was 54 $\pm$ 3\% (90\% confidence level). The pulsed fractions of the hard and soft bands derived using the same method were 56 $\pm$ 3\% and 62 $\pm$ 2\% respectively. In addition, we derive an upper limit of $\sim$50$\%$ for the pulsed fraction in the \emph{ASCA} data (90$\%$ confidence level). The lack of detection of the 626 s period in this data could therefore be due to a slight drop in the pulsed fraction, although it is also possible that the modulation was not on during the earlier observation.

Using $\tt{XSPEC}$\footnote{http://heasarc.nasa.gov/docs/xanadu/xspec/} v12.5.0 we fitted a number of spectral models common to CVs and X-ray binaries to the MOS
spectra (Table \ref{models}). We first attempted simple models described by absorbed power law, bremsstrahlung, and thermal plasma (MEKAL) models. Of these, only the power law provided an acceptable fit. The temperatures for the bremsstrahlung and MEKAL models could not be constrained, and neither could the metal abundances for the MEKAL fit ($\sim$15 times Solar). The fit residuals for these two models indicated problems arising in the low-energy region of the spectrum around 1 keV. The addition of a blackbody (BB) component with kT $\sim$3 keV improved the fits significantly, although the plasma and bremsstrahlung temperatures and the MEKAL metal abundance ($\sim$3 times Solar) were poorly constrained. The final fit we attempted was a multi-temperature thermal plasma model (CEVMKL), with a maximum temperature of kT $\sim$ 100 keV and dramatically non-Solar abundances. However, an acceptable fit could not be reached with this model, and none of the model parameters could be constrained (hence approximate values only are given in Table \ref{models}). We therefore conclude that the model which best physically represents the observed spectrum is an absorbed power law. 


Table \ref{specpar} presents
the parameters for the adopted absorbed power law spectral model. The best-fit spectral parameters are consistent with those of the \emph{ASCA} source AX J1740.2$-$2903 \citep[n$_H$ = 0.1$^{+1.0}_{-0.1}$ $\times$ 10$^{22}$ atom cm$^{-2}$, $\Gamma$ = 0.6$^{+0.6}_{-0.4}$;][]{sak02}, which had a 0.2 -- 10 keV unabsorbed flux of F$_X$ $\sim$ 5  $\times$ 10$^{-12}$ erg s$^{-1}$ cm$^{-2}$, approximately consistent with that of \puls. Using the same spectral parameters in Table \ref{specpar} and the $\emph{ROSAT}$ PSPC 0.1 -- 2.4 keV count rate of 0.019 $\pm$ 0.003 count s$^{-1}$ \citep{sid01}, the unabsorbed 0.2 -- 10 keV flux of the ROSAT source AX J1740.3$-$2904 is (5.6 $\pm$ 0.7) $\times$ 10$^{-12}$ erg s$^{-1}$ cm$^{-2}$. This value (which does not take into account the uncertainties of the spectral parameters) can thus be taken as consistent with the EPIC flux of \puls\ within the uncertainties of the method.

In order to further investigate the possibility of spectral variability over the 626 s period, we extracted
spectra from the high (phases 0.3 -- 0.7 in Figure \ref{tfold}) and low (phases 0.8 -- 1.2 in Figure \ref{tfold}) states separately and fitted them with the same absorbed power law model.
The best-fit spectral parameters obtained for each set of spectra are listed in Table \ref{specpar}. Neither the column density nor the power law photon index vary within the derived uncertainties, indicating that the shape of the continuum spectrum does not appear to change significantly between the high and low states within the limits of the statistics of the data. 

\begin{table*}
\begin{center}
\caption[]{Spectral models applied to the MOS1 and
MOS2 spectra. Column 3 gives the photon index or plasma temperature of the first component. Column 4 gives the temperature of the black body component when applicable. For the CEVMKL model, column 3 gives the power law index for the emissivity function, and column 4 gives the maximum plasma temperature kT$_{max}$. The fluxes quoted are unabsorbed in the range 0.2 -- 10 keV. Approximate values are given where parameters could not be constrained. Quoted errors indicate the 90$\%$ confidence levels.}\label{models}
\begin{tabular}{lccccc}
\hline Model & n$_H$ & $\Gamma$/kT$_1$ & kT$_2$ & Flux &  $\chi^2$/d.o.f\\
 & (10$^{22}$ atom cm$^{-2}$) & ( -- / keV) & (keV) & (10$^{-12}$ erg cm$^{-2}$ s$^{-1}$)& \\
\hline 
MEKAL + BB & 0.6$^{+0.2}_{-0.2}$ & 10$^{+9}_{-4}$ & 2.7$^{+1.0}_{-0.6}$ & 4.0$^{+0.4}_{-0.3}$&  80.3/85\\
Power law & 0.4$^{+0.2}_{-0.1}$&0.5$^{+0.1}_{-0.1}$&--&4.0$^{+0.3}_{-0.2}$& 87.6/89 \\
Bremsstrahlung + BB & 0.5$^{+0.3}_{-0.2}$ & 10$^{+10}_{-10}$&3.0$^{+2.0}_{-0.7}$& 4.1$^{+0.3}_{-0.2}$&86.1/87 \\
MEKAL & 1.1$^{+0.2}_{-0.2}$&$\sim$80&--&4.1$^{+0.3}_{-0.3}$& 133.3/87\\
Bremsstrahlung & 1.1$^{+0.2}_{-0.2}$& $\sim$200&--&3.9$^{+0.3}_{-0.3}$& 136.5/89 \\
CEVMKL &$\sim$1 &$\sim$0.8 &$\sim$100 &$\sim$5 & 119.6/74\\
\hline
\end{tabular}
\end{center}
\end{table*}

   \begin{figure}
   \begin{center}
   \includegraphics[angle=-90,width=\columnwidth]{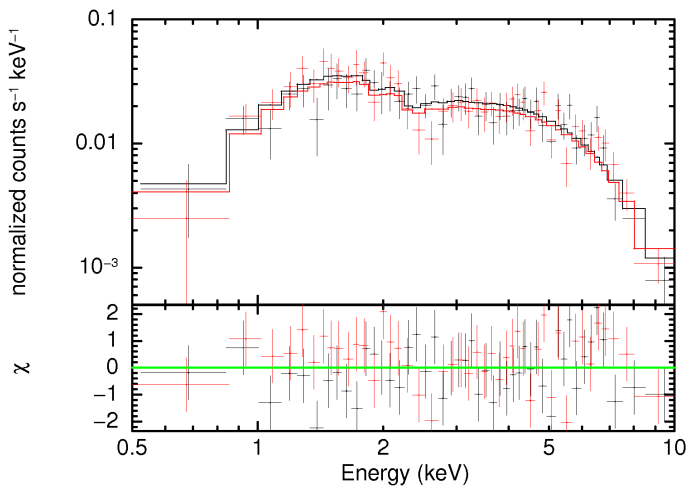}
   \caption{MOS1 (black) and MOS2 (red) spectra for all data fitted
     with an absorbed power law model. The $\chi^2$ residuals to the fit are shown in the bottom panel.}\label{spec}
    \end{center}
    \end{figure}

   \begin{figure}
   \begin{center}
   \includegraphics[angle=-90,width=\columnwidth]{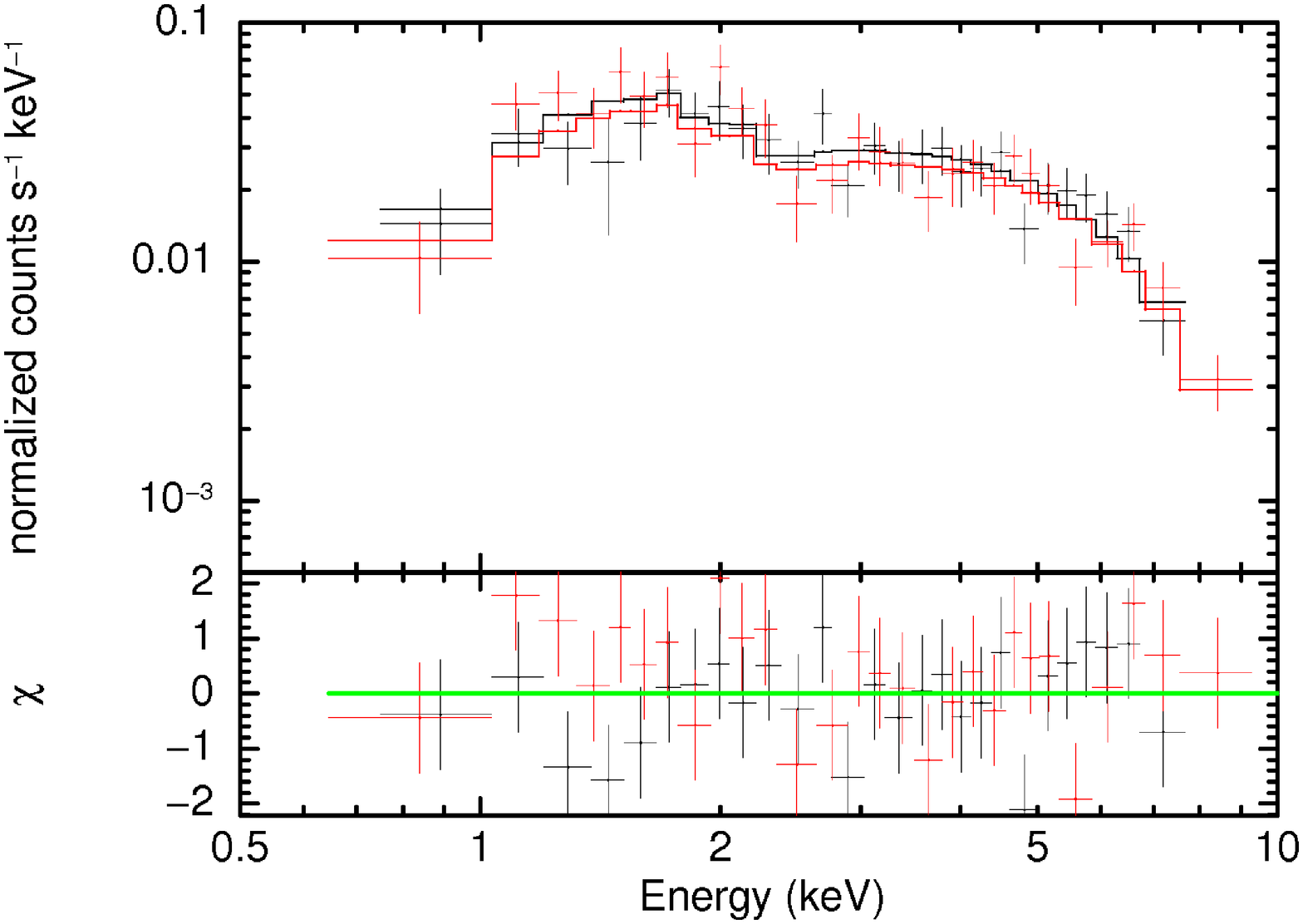}
   \caption{MOS1 (black) and MOS2 (red) spectra for the high state
     fit using an absorbed power law model.}\label{hspec}
    \end{center}
    \end{figure}
    
       \begin{figure}
   \begin{center}
   \includegraphics[angle=-90,width=\columnwidth]{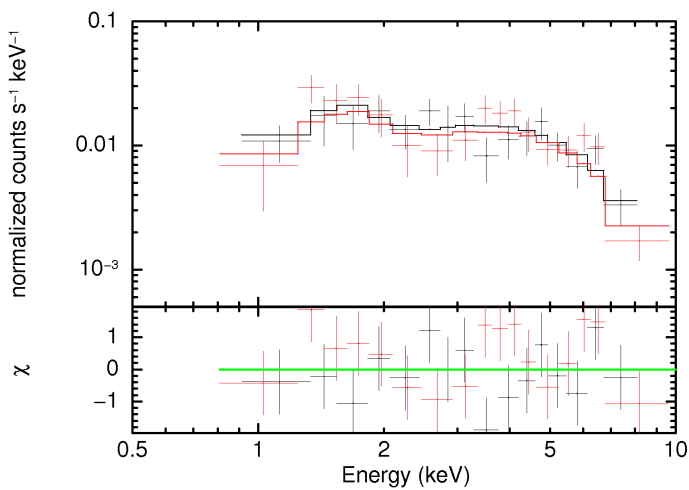}
   \caption{MOS1 (black) and MOS2 (red) spectra for the low state
     fit using an absorbed power law model.}\label{lspec}
    \end{center}
    \end{figure}

\section{Candidate Optical/NIR Counterparts}

Numerous optical and NIR databases were searched for imaging data
covering the field of \puls. No match was found within the \xmm\ error circles in the 2MASS All-Sky Point Source Catalogue \citep{cut03}. However, there was a faint optical counterpart detected in the USNO-B1.0 catalogue \citep{mon03}, 1.4$\arcsec$ offset from the \xmm\ position and well within the error circle. The magnitudes of this optical counterpart are given as $R1$ = 18.04 mag, $B2$ = 19.07 mag, $R2$ = 18.78 mag, and $I$ = 17.44 mag. There was no detection reported in the $B1$-band. These magnitudes are close to the detection limit in the USNO-B1.0 catalogue (especially the $I$-band detection) and therefore we cannot say anything meaningful about the nature of this object based on the optical photometry.

The best imaging data were found in the \dr\ (United Kingdom Infrared Deep Sky Survey --
Galactic Plane Survey: Data Release 5) database. UKIDSS is a
\nir\ survey covering approximately 7000 deg$^2$ of the Northern
hemisphere sky to a depth of $K = 18$ mag, with additional data from
two deeper, small area high-redshift galaxy surveys.  Using the Wide
Field Camera (WFCAM) on the United Kingdom Infrared Telescope (UKIRT),
the survey achieved a pixel resolution of $0.14\arcsec$ by use of the
micro-stepping technique \citep[see][for full details]{lawr07}. The
data used in this paper was taken from the UKIDSS-GPS, a survey of
approximately 2000 deg$^2$ of the Northern Galactic plane in the $J$,
$H$ and $K$-bands \citep{luca07}.

\begin{table}[h!]
\begin{center}
\caption[]{Best-fit spectral parameters for the adopted model fitted to the combined
(total), high state and low state (phases 0.3 -- 0.7 and phases 0.8 -- 1.2 respectively in Figure \ref{tfold}) spectra. The fluxes quoted are unabsorbed in the 0.2 -- 10 keV range. Quoted errors indicate the 90$\%$ confidence levels.}\label{specpar}
\begin{tabular}{lcccc}
\hline Param. & Total & High & Low & Units\\
\hline 
$n_H$& 0.4 $^{+ 0.2}_{- 0.1}$ & 0.4 $^{+ 0.2}_{- 0.2}$ &  0.4 $^{+ 0.5}_{- 0.3}$&10$^{22}$ atom cm$^{-2}$\\
$\Gamma$ & 0.5 $^{+ 0.1}_{- 0.1}$& 0.6$^{+ 0.2}_{- 0.2}$ & 0.3 $^{+ 0.3}_{- 0.2}$& --\\
Norm. & 1.3 $^{+ 0.3}_{- 0.2}$ & 1.8 $^{+ 0.5}_{- 0.5}$ & 0.7 $^{+ 0.5}_{- 0.2}$& 10$^{-4}$\\
Flux & 4.0 $^{+ 0.3}_{- 0.2}$ & 5.3 $^{+ 0.4}_{- 0.4}$& 3.0 $^{+ 0.4}_{- 0.3}$ &10$^{-12}$ erg cm$^{-2}$ s$^{-1}$\\
$\chi^2$/dof & 87.6/89 & 47.1/53 &28.9/30&--\\
\hline
\end{tabular}
\end{center}
\end{table}

\begin{figure}[b!]
  \begin{center}
    \includegraphics[width=\columnwidth]{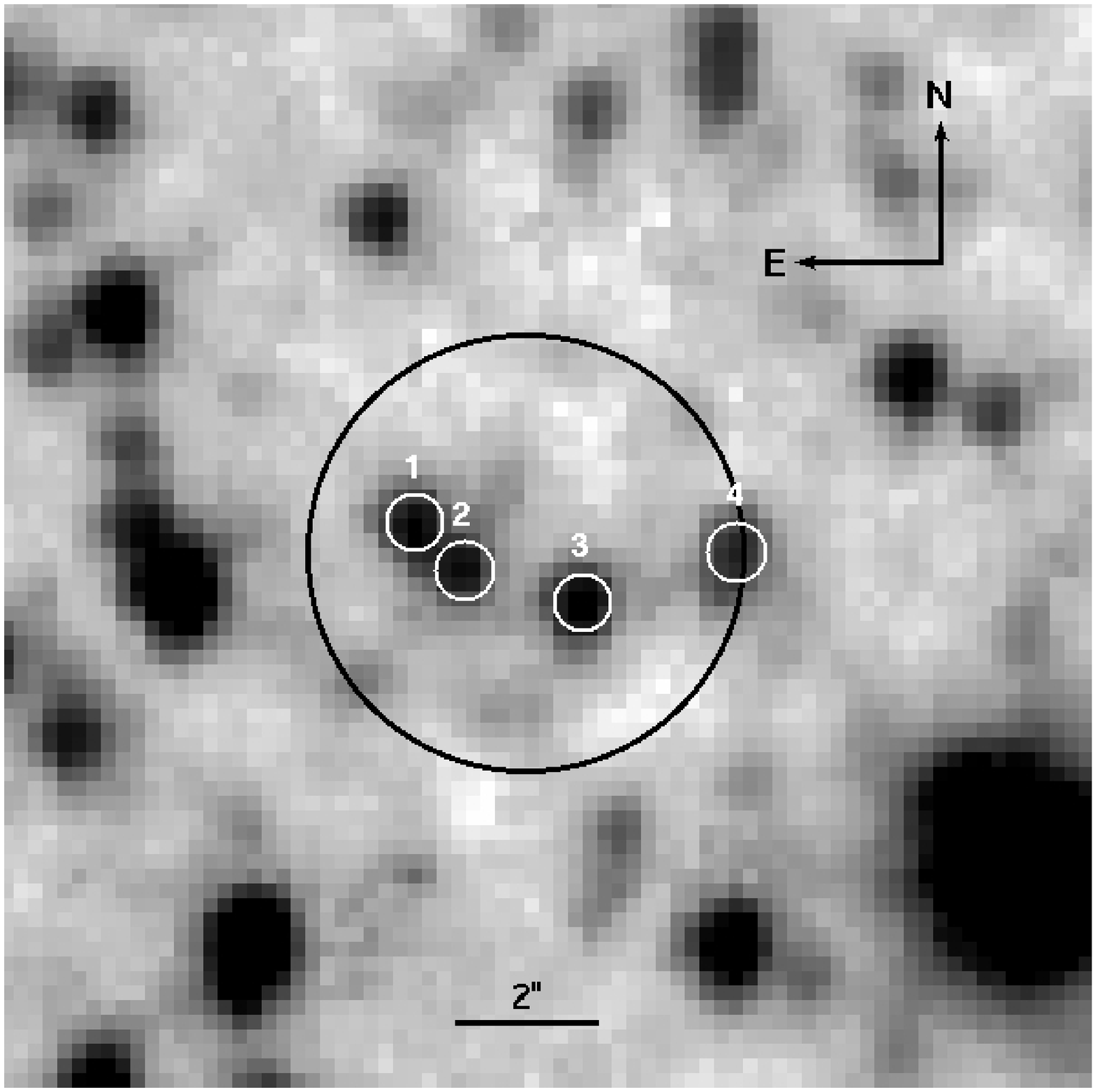}
    \caption{$15\arcsec \times 15\arcsec$ $\it{JHK}$ finding chart
      for \puls . The large black circle is centred on the
      \xmm\ position of \puls, with the radius indicating the
      3\arcsec\ positional error. The white circles represent the four
      candidate counterparts, hereafter referred to as sources 1, 2, 3
      and 4 from left to right respectively. Sources 1 and 2 were
      classed as a single source in the \dr\ database prompting the
      manual re-extraction of source astrometry and photometry.}
    \label{f:finder}
  \end{center}
\end{figure}

Three NIR sources were identified in the \dr\ database that were
astrometrically coincident with the 3\arcsec\ error circle of \puls.
Investigation of the UKIDSS images found that one of these sources was
in fact two blended sources that the UKIDSS pipeline had been unable
to separate (sources 1 and 2 in Figure \ref{f:finder} and Table
\ref{t:positions}). These two blended sources are conicident
  with the faint detection in the USNO-B1.0 catalogue indicating that the faint optical
  source is likely to be associated with either or both of these
 blended sources.

\subsection{Data Reduction}

Due to the crowding in the field, manual extraction of the
astrometry and photometry was required to separate the blended
source (sources 1 and 2) within the \xmm\ error circle. The
\sext\footnote{\url{http://terapix.iap.fr/}} software package was
utilised to extract the astrometry and photometry of the sources
from the UKIDSS images. The default settings were employed for all
but two of the \sext\ input variables. In order to maximise the
likelihood of resolving blended sources, the number of de-blending
thresholds was increased to the maximum value of 64. This was
successful in resolving the two blended sources in the $K$-band
image, but not in the $J$- or $H$-bands. In order to resolve the
blended sources we applied a Mexican Hat convolution filter to the
images, successfully allowing the program to resolve the blended
source in all three bands. \sext\ thus extracted astrometry and
photometry for all sources within a 1\arcmin\ box surrounding
\puls.

\begin{table}
  \begin{center}
    \caption{Positions of the 4 candidate NIR
      counterparts to \puls. The 3$\sigma$ positional errors are 0.2$\arcsec$.}
    \begin{tabular}{ccc}
      \hline
      Star & R.A. & Dec. \\
      \hline
      1 & 17h 40$\arcmin$ 16.15$\arcsec$ & -29$\degr$ 03$\arcmin$ 37.4$\arcsec$ \\
      2 & 17h 40$\arcmin$ 16.09$\arcsec$ & -29$\degr$ 03$\arcmin$ 38.0$\arcsec$  \\
      3 & 17h 40$\arcmin$ 15.97$\arcsec$ & -29$\degr$ 03$\arcmin$ 38.5$\arcsec$ \\
      4 & 17h 40$\arcmin$ 15.81$\arcsec$ & -29$\degr$ 03$\arcmin$ 37.8$\arcsec$ \\
      \hline
    \end{tabular}
    \label{t:positions}
  \end{center}
\end{table}

Once the astrometry and photometry were extracted for the sources in
all three bands, the source lists were combined to form a single
catalogue listing the mean astrometric position, magnitude, and
colours for each source. Comparison of this catalogue to that obtained
from the WSA archive showed that the astrometry derived by \sext\ was
within 0.05\arcsec\ of that derived by the UKIDSS pipeline. The
photometry of the sources matched that of the UKIDSS catalogue at the
0.05 mag level, with the errors produced by the \sext\ package having
the same distribution as those in the archive. The errors quoted for
all sources include the additional 0.05 mag error accounting for the
difference between the \sext\ and UKIDSS magnitudes (see Table
\ref{t:redcol}).

\begin{table*}
  \begin{center}
    \caption{Reddened magnitudes and colours of the 4 candidate
      counterparts to \puls. The colours of source 2 correspond to a
      foreground M\,5/6~V star whose magnitudes indicate a distance of
      $\sim$520\,pc. The colours and magnitudes of the remaining 3
      sources are inconsistent with any type of stellar source
      indicating they are heavily reddened, most likely in the
      Galactic Bulge (see Figure \ref{f:candscmd}).}
    \begin{tabular}{ccccccccc}
      \hline
      \hline
      Star & $J$ & $H$ & $K$ & \jh & \jk & \hk & Spectral Type & Distance \\
      & (mag)& (mag)& (mag)& (mag)& (mag)& (mag)&&(kpc)\\
      \hline
      1 & $16.38 \pm 0.12$ & $15.07 \pm 0.12$ & $14.50 \pm 0.12$ & $1.30 \pm 0.17$ & $1.88 \pm 0.17$ & $0.57 \pm 0.17$ & --- & --- \\
      2 & $15.71 \pm 0.06$ & $15.01 \pm 0.07$ & $14.70 \pm 0.08$ & $0.70 \pm 0.09$ & $1.02 \pm 0.10$ & $0.31 \pm 0.11$ & M\,5/6~V & 0.5\\
      3 & $16.62 \pm 0.08$ & $15.21 \pm 0.07$ & $14.48 \pm 0.07$ & $1.41 \pm 0.11$ & $2.15 \pm 0.11$ & $0.73 \pm 0.10$ & --- & --- \\
      4 & $16.93 \pm 0.09$ & $15.70 \pm 0.09$ & $14.98 \pm 0.08$ & $1.22 \pm 0.13$ & $1.95 \pm 0.12$ & $0.72 \pm 0.12$ & --- & --- \\
     \hline
      \hline
    \end{tabular}
    \label{t:redcol}
  \end{center}
\end{table*}

\subsection{The Extinction}

As shown in \citet{gosl07} and \citet{gosl09}, the extinction towards
the Nuclear Bulge is extremely complex. The extinction measures
derived from earlier studies are insufficient to obtain accurate
photometry of Bulge sources. Using the extinction measured by
\citet{dutr03} and the ``standard'' conversion factors of
\citet{riek85}, we cannot recover consistent photometry for a single
spectroscopic type of star for most of the candidate
counterparts. Therefore, we have used the techniques described in
\citet{gosl09} to correct for the extinction.  Median colours and
magnitudes were calculated for all stars within 20\arcsec\ for which
there was photometry in the $J$-, $H$- and $K$-bands (see Figure
\ref{f:candscmd}). Colour excesses were then obtained by subtracting
the median intrinsic colour values for all giant type stars in the
UKIRT \nir\ standards
catalogue\footnote{\url{http://www.jach.hawaii.edu/UKIRT/astronomy/calib/phot_cal/}}.
The resulting colour excesses corresponded to an average extinction
law value of $\alpha$ = 1.365, and absolute extinction values of $A_J
= 2.875 \pm 0.07\,{\rm mag}$, $A_H = 2.015 \pm 0.1\,{\rm mag}$ and
$A_K = 1.405 \pm 0.05\,{\rm mag}$. These values correspond to
  $A_V \ge 15\,{\rm mag}$ meaning that the faint optical counterpart
  detected in the USNO-B1.0 catalogue must be a relatively local source.
Correcting the magnitudes of the candidate counterparts gave the
colours and magnitudes shown in Table \ref{t:unredcol}.

\begin{table*}
  \begin{center}
    \caption{De-reddened magnitudes and colours of the 4 candidate
      counterparts to \puls. The colours and magnitudes of source 2
      have been de-reddened to show that an extinction correction does
      not lead to colours or magnitudes that correspond to a single
      spectral type. The colours of the stars after extinction
      correction are used to identify probable spectral types and
      magnitudes then used to estimate the distance to such a star for
      the other 3 candidate counterparts.  There are two possible
      solutions to the spectral type of source 3, one a distant giant,
      the second a more local dwarf.  The position of source 3 on the
      colour-magnitude diagram (Figure \ref{f:candscmd}) suggest the
      distant giant is the more likely real spectral type.}
    \begin{tabular}{ccccr@{.}lr@{.}lr@{.}lcc}
      \hline
      \hline
      Star & $J$ & $H$ & $K$ & \multicolumn{2}{c}{\jh} & \multicolumn{2}{c}{\jk} & \multicolumn{2}{c}{\hk} & Spectral Type & Distance \\
    & (mag)& (mag)& (mag)& \multicolumn{2}{c}{(mag)}& \multicolumn{2}{c}{(mag)}& \multicolumn{2}{c}{(mag)}&&(kpc)\\
      \hline
      1 & $13.50 \pm 0.14$ & $13.05 \pm 0.16$ & $13.09 \pm 0.13$ & $ 0$&$44 \pm 0.21$ & $ 0$&$41 \pm 0.19$ & $-0$&$04 \pm 0.21$ & G/K~V & 0.5--1.1\\
        & --- \arcsec\ --- & --- \arcsec\ --- & --- \arcsec\ --- & \multicolumn{2}{c}{--- \arcsec\ --- } & \multicolumn{2}{c}{--- \arcsec\ --- } & \multicolumn{2}{c}{--- \arcsec\ --- } & K1~III & 8.3\\
      2 & $12.83 \pm 0.09$ & $12.99 \pm 0.12$ & $13.29 \pm 0.09$ & $-0$&$16 \pm 0.15$ & $-0$&$56 \pm 0.13$ & $-0$&$30 \pm 0.15$ &   ---   &   ---   \\
      3 & $13.74 \pm 0.11$ & $13.19 \pm 0.12$ & $13.07 \pm 0.09$ & $ 0$&$55 \pm 0.16$ & $ 0$&$67 \pm 0.14$ & $ 0$&$12 \pm 0.15$ & K4~V &0.5\\
        & --- \arcsec\ --- & --- \arcsec\ --- & --- \arcsec\ --- & \multicolumn{2}{c}{--- \arcsec\ ---} & \multicolumn{2}{c}{--- \arcsec\ ---} & \multicolumn{2}{c}{--- \arcsec\ ---} & K1~III & 8.7\\
      4 & $14.05 \pm 0.11$ & $13.68 \pm 0.13$ & $13.57 \pm 0.09$ & $ 0$&$37 \pm 0.17$ & $ 0$&$48 \pm 0.14$ & $ 0$&$11 \pm 0.16$ & G5--K5~V & 0.6--1.3\\
      \hline
      \hline
    \end{tabular}
    \label{t:unredcol}
  \end{center}
\end{table*}

\subsection{Candidate Counterpart Identification}

We compared the corrected colours and magnitudes of the candidate
counterparts to the standards in the UKIRT standards catalogue to
identify the spectroscopic type. The de-reddened photometry of source
1 indicates that it is either a G or K type dwarf at $\sim$0.5 or
$\sim$1 kpc respectively, or at the limits of the errors given by the
source extraction, anything earlier than a K1~III at a distance of up
to 8\,kpc. The photometry for this source does not provide a good
match to any single spectral type (or small range of types), so it is
possible that the photometry is being affected by residuals from the
deblending with the nearby source 2. It is also possible that
  the accretion disc in the \puls\ system (if source 1 is the true counterpart) is contributing to the
  observed photometry which is why it is not possible to obtain a firm
  identification of the spectral type using this method.  Further,
  higher resolution images or spectroscopy will be necessary to
  resolve this source for further study as a candidate counterpart to
  \puls.~Spectroscopy would also allow us to identify any accretion
  disc contribution to the source.

\begin{figure*}
  \begin{center}
    \includegraphics[width=0.66\columnwidth]{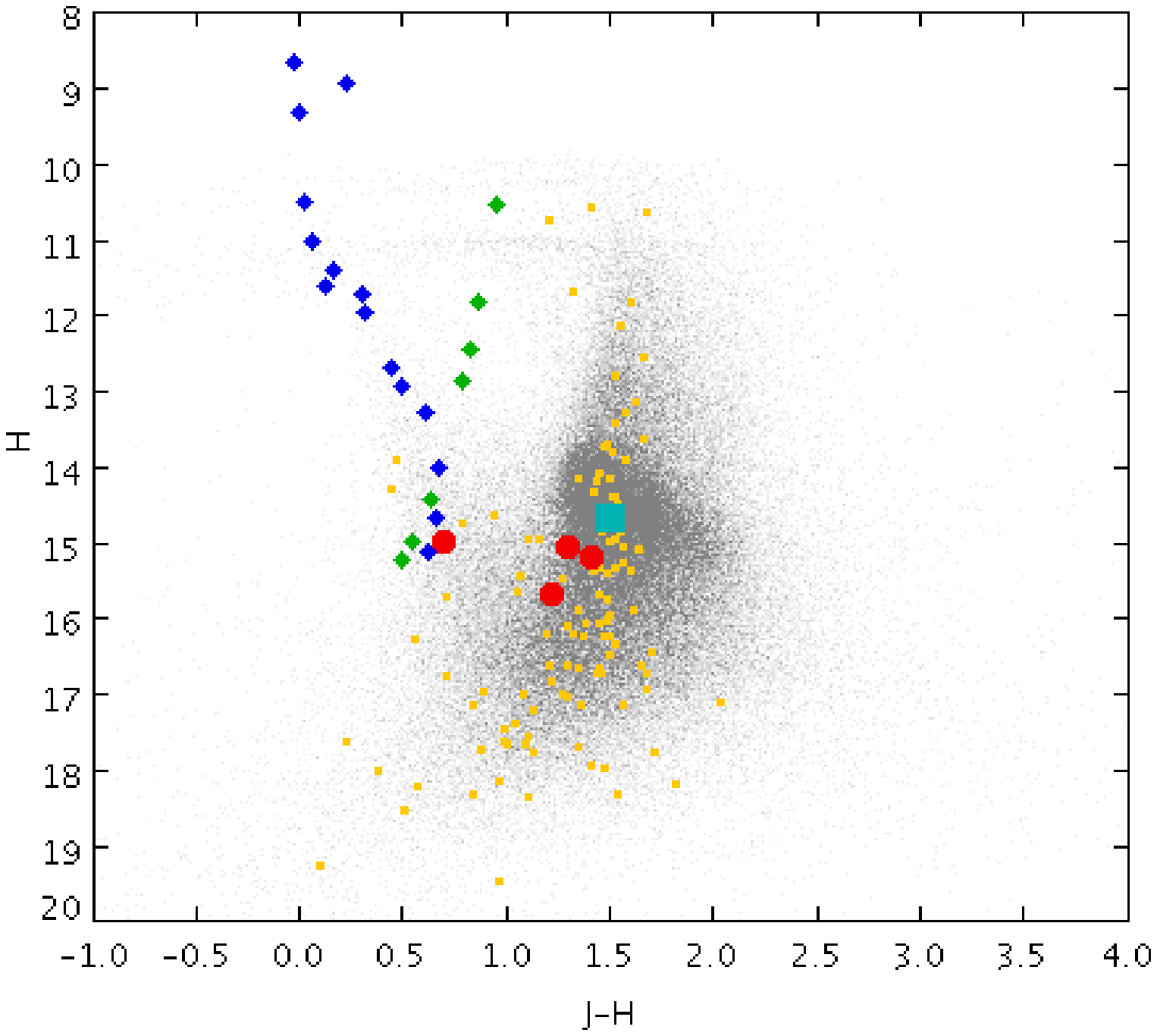}
    \includegraphics[width=0.66\columnwidth]{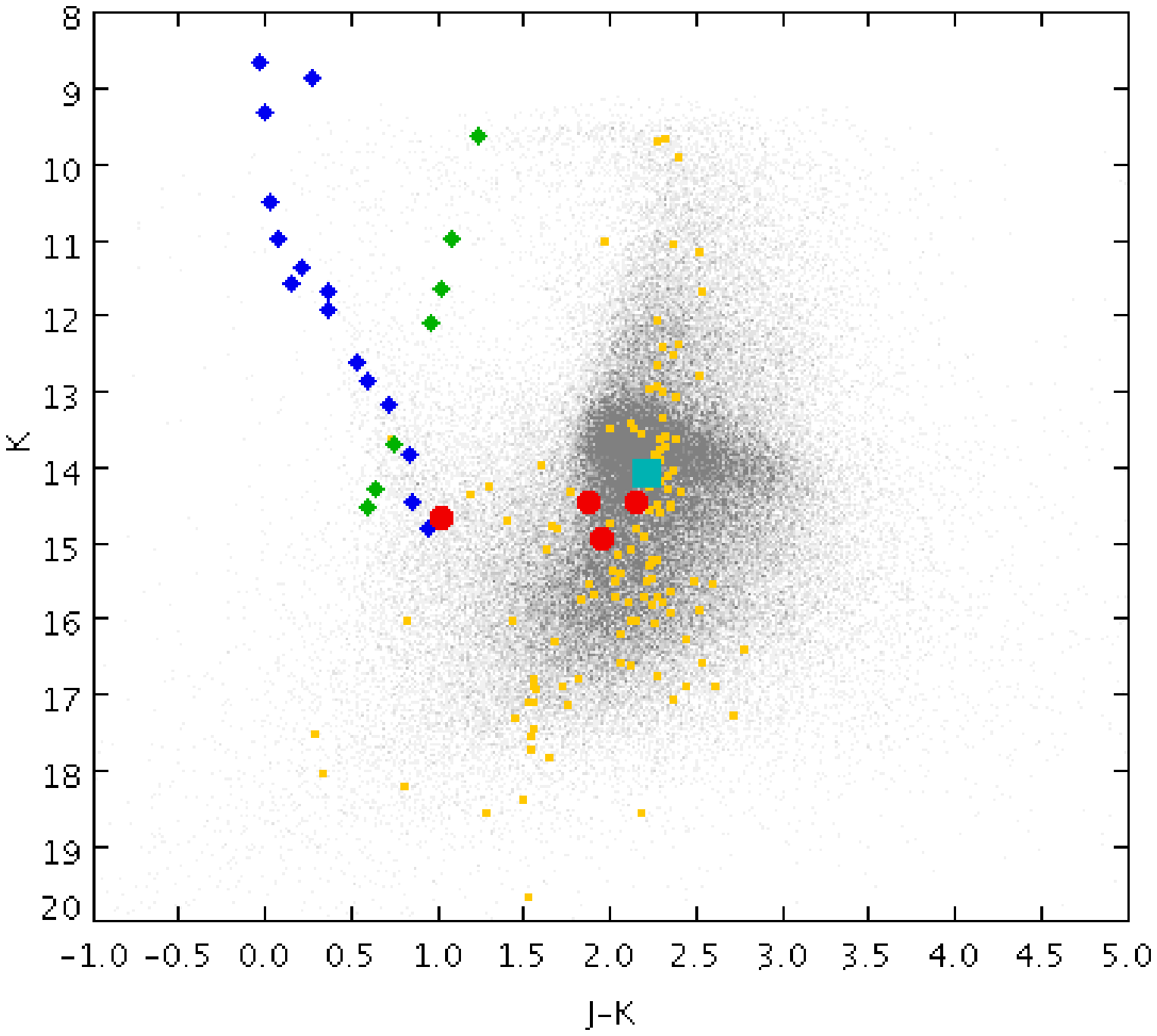}
    \includegraphics[width=0.66\columnwidth]{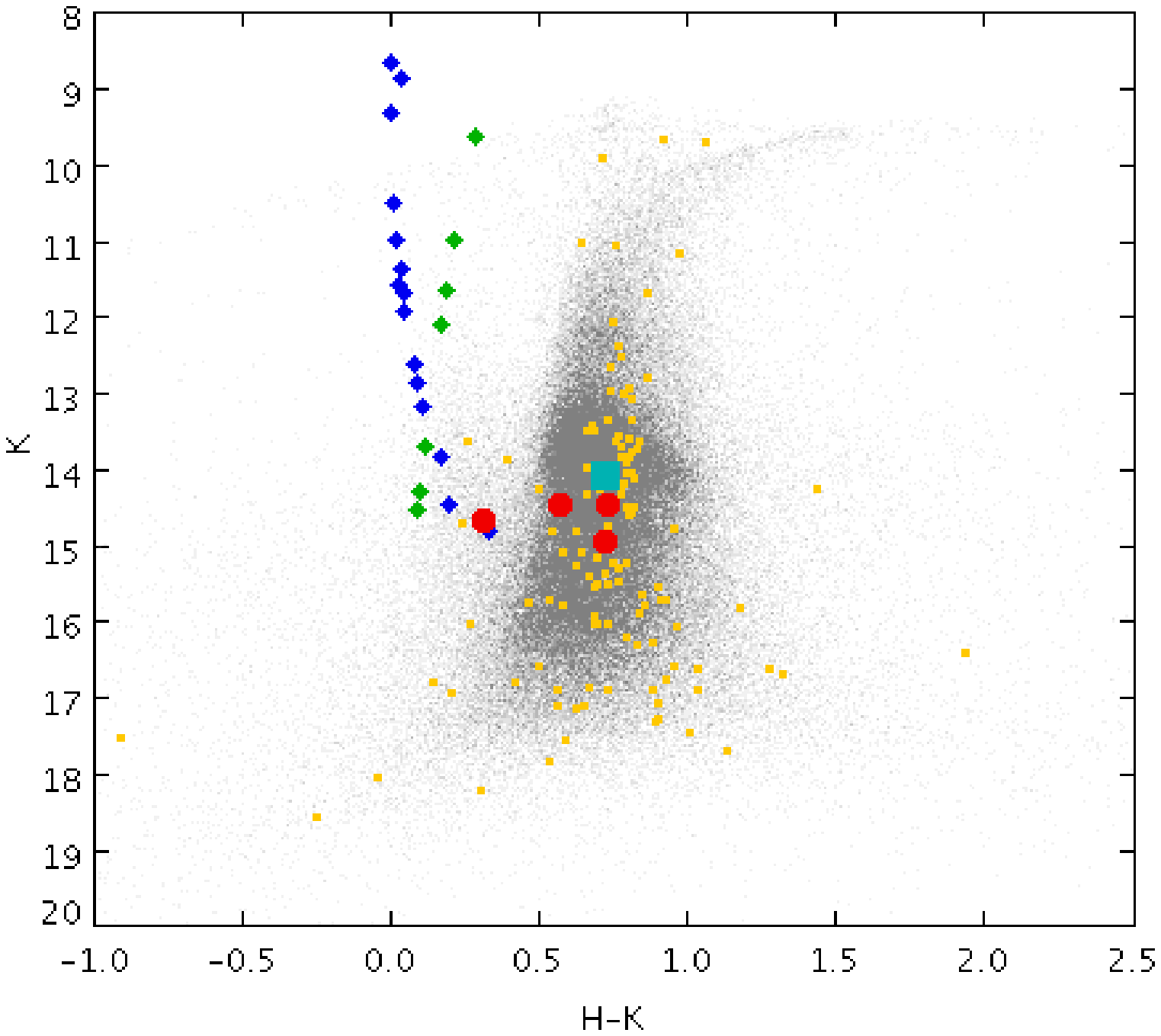}
    \caption{Colour magnitude diagrams of the candidate counterparts
      to \puls\ and the surrounding field stars. The grey points are
      the field stars within 3\arcmin\ of the X-ray source.  The
      majority of this field population are Bulge giants forming the
      main concentration of points in the centre of each diagram, but
      there is also a small local dwarf population with lower colour
      values than the giant population.  The 4 red circles represent
      the 4 candidate counterparts, with sources 2, 1 and 3 (from left
      to right respectively) all at approximately the same magnitude,
      and source 4 being the slightly fainter object in all three
      diagrams (see text and tables for details of the source
      numbering). The yellow points are the stars within 20\arcsec\ of
      \puls\ used for the extinction calculation and the light-blue
      square is the median value of these stars colours and
      magnitudes. The dark blue diamonds are main sequence standard
      stars taken from the UKIRT standard catalogue corrected for a
      distance modulus corresponding to 550\,pc. These trace the
      observed local dwarfs in the field well and one of the candidate
      counterparts (source 2) has colours and magnitudes consistent
      with the latest M type dwarf. The green diamonds are giant
      standards also from the UKIRT standard catalogue corrected for a
      distance modulus corresponding to a distance of 8\,kpc and with
      the measured extinction of 2.0\,mag and 1.4\,mag added to the
      $H$ and $K$-bands respectively.}
    \label{f:candscmd}
  \end{center}
\end{figure*}

The colours and magnitudes of source 2 when corrected for extinction
do not match to any spectral type for either dwarf, giant of
supergiant stars (see Table \ref{t:unredcol}). The position of this
source in all three colour-magnitude diagrams in Figure
\ref{f:candscmd} suggests that it is actually a relatively local dwarf
and not a Bulge giant. Before extinction correction, comparison of the
colours and magnitudes of this star to standards corresponds to an
M5/6~V star at a distance of $\sim$0.5\,kpc (see Table
\ref{t:redcol}).  It is likely that this local source is the
  NIR counterpart of the faint optical source detected in the USNO-B1.0 catalogue.

Source 3 has de-reddened colours and magnitudes consistent with either
a K4~V star at a distance of $\sim$0.5\,kpc, or a K1~III at a distance
of $\sim$8.7\,kpc (placing it in the Bulge). Source 3 is the candidate
counterpart with the highest colour values, and it is difficult to
explain how a local dwarf would require such a large extinction
correction. We therefore conclude that the most likely spectral type
for this star is that of a K1~III.

Source 4, which appears to be slightly extended, has de-reddened
colours that are consistent with a G5 -- K5~V star at a distance of
$\sim$0.6 -- 1.3\,kpc (see Table \ref{t:unredcol}). The colours are
too blue to correspond to a giant star except at the extremes of the
errors. This source is the faintest of the candidate counterparts,
being the ``lowest'' source (marked in red) in all three
colour-magnitude diagrams in Figure \ref{f:candscmd}. Although its
position in the colour-magnitude diagrams suggests that it is part of
the Bulge giant population, it is also possible that it is a less
distant dwarf as there is a blending of the two populations at low
magnitudes towards the bottom of the colour-magnitude diagrams. As the
source appears extended, an alternative explanation is that it is a
blend of more than one source that the \sext\ routine was unable to
resolve, or a background galaxy. However, a location beyond the
Galactic bulge is not consistent with the neutral hydrogen column
density derived from the X-ray spectral fitting, so it is unlikely
that \puls\ is a distant object. As for source 1, it is also
  possible that the photometry of source 4 is affected by a contribution
  from a disc, which would result in our inability
  to identify the spectroscopic type based solely on the
  photometry.  As such, higher resolution imaging of this source and
  preferably spectroscopy will be necessary in order to better
  determine the spectral type of this candidate counterpart and
  identify any disc contribution.

\section{Discussion} 

The 626 s period could be interpreted as either a modulation of the X-ray emission over the orbit of a compact binary system, or the spin of a compact object (in this case either a white dwarf or neutron star). The hard X-ray spectrum and long period of \puls\ rule out an isolated rotation powered pulsar, as the synchrotron spectra of these objects tend to have a steeper spectral slope and typically have spin periods $<$ 1s \citep[see e.g.][]{li08}. 

A 626 s period is consistent with either the spin of a white dwarf in an IP or the orbit of a double white dwarf AM CVn system \citep[e.g.][]{ku06}. The X-ray spectra of IPs can generally be represented by a hard bremsstrahlung continuum with a temperature of a few tens of keV with the addition of various emission lines (including the Fe K$\alpha$ lines), sometimes with the inclusion of a soft thermal excess commonly represented by a BB component with kT $<$ 0.5 keV \citep{ku06}. Attempts to fit the spectra of \puls\ with absorbed bremsstrahlung or thermal plasma (MEKAL) models required the addition of a BB component to represent the soft excess. Although acceptable fits were obtained, the component temperatures were poorly constrained, and the BB temperatures were outside the physical range for an IP. 

The optical emission of IPs is typically dominated by emission from the hot accretion column and truncated accretion disc, and should emit strongly in the UV band. However, the lack of detection in the OM \emph{uvw2} image (with a derived 3$\sigma$ upper limit of 20.3 mag) does not allow us to place any constraints on the nature of this object, due to uncertainty in the level of extinction in the UV bands. \citet{hal10a} obtained optical spectra of the USNO B-1.0 counterpart in the wavelength range 371 -- 738 nm, detecting emission lines consistent with the H$\alpha$, H$\beta$, HeI, and HeII 4686 \AA~lines at zero redshift. They concluded that the relative strengths of these lines, the moderate reddening, the X-ray period and the X-ray spectrum suggested an IP nature.

IPs generally have 2 -- 10 keV X-ray luminosities of less than about 10$^{33}$ erg s$^{-1}$ \citep{war95}, which would put \puls\ at a distance of $<$ 2 kpc. This is consistent with each of the candidate NIR candidates and the optical counterpart, supporting the possibility that \puls\ could be an IP. In addition, the optical photometric period of 622 $\pm$ 7 s reported by \citet{hal10b} (if confirmed) would support the association of \puls\ with the USNO-B1.0 optical counterpart, and would argue strongly against a location near the Galactic bulge. However, IPs typically show absorption dips in their light curves over the spin periods, caused by the passage of accretion curtains across the line of sight \citep{eva07}. In IP systems that do not show absorption dips (thought to be due to geometric viewing angles), a low-temperature (kT $\lesssim$ 100 eV) BB component is typically found in their X-ray spectra \citep{eva07}. In addition, IPs typically show strong emission lines in their X-ray spectra associated with the Fe K$\alpha$ complex \citep{hel04}. None of these features (i.e. absorption dips, low-temperature BB component, or Fe K$\alpha$ lines\footnote{\citet{mal10} reported the presence of an Fe line in their analyses of the same EPIC data presented here. Our initial cursory analysis of the spectra produced by the \xmm\ pipeline support the presence of an emission feature around 6.7 keV. However, careful manual reduction of the EPIC data using the latest calibration files and correct response matrices found no evidence of such a feature.}) were found in the EPIC data, although this may be due to inadequate statistics. We therefore cannot rule out that \puls\ is an IP, although its properties are unlike any other IP that has been studied.

A nearby AM CVn would not be expected to have a bright optical counterpart. The X-ray spectra of these systems are more complex, requiring multi-temperature thermal plasma models in which the emission measures follow a power-law distribution, and significantly non-Solar elemental abundances \citep[the CEVMKL model in $\tt{XSPEC}$;][]{ra05}. Attempts to fit the spectra of \puls\ with a similar model could not achieve an acceptable fit, and obtained a maximum temperature of $>$ 100 keV, far in excess of the temperatures observed from other AM CVns \citep{ra05}. We therefore conclude that \puls\ is unlikely to be an AM CVn system.  
 
The hard power law X-ray spectrum of \puls\ is consistent with emission produced from accretion onto a magnetised neutron star \citep[see e.g.][]{nag89}. In this scenario the 626 s period could represent either an UCXB orbital period  \citep[see e.g.][]{int07} or the spin of a slow accretion powered pulsar  \citep[see e.g.][]{ikh07}. As described in $\S$1, globular clusters are host to five times more UCXBs than the field. However, a Galactic bulge location for \puls\ could also provide a dense environment where a 626 s orbit UCXB might form. The shortest ultra-compact orbital period belongs to the globular cluster source 4U 1820-303 with an orbital period of 685 s \citep{whi86}. Therefore if the 626 s period in \puls\ is orbital in nature, it would be the shortest period UCXB currently known.

X-ray orbital modulations in ultra-compact binaries are typically associated with eclipses or dips in high inclination systems. However, the sinusoidal profile (Figure \ref{tfold}) and lack of discernible absorption variability over the 626 s period argue against full or partial eclipses by the disc or donor star. Orbital modulation in X-rays could also be produced by variable accretion in a system with an eccentric orbit, although the orbits of UCXBs should circularise over very short timescales due to tidal interactions. However, the presence of a third body in a hierarchical triple system may tidally induce a non-zero eccentricity in the inner binary. The globular cluster UCXB 4U 1820-303 has been proposed as such a system, where the 685 s inner neutron star/white dwarf binary orbit is thought to be accompanied by a $\sim$1.1 d orbit companion \citep{ch01,zdz07}. In this system the presence of the third body is derived via the detection of a longer-term $\sim$170 d periodic modulation in X-rays, thought to be linked to oscillations of the inner binary eccentricity caused by tidal interactions with the third body.  However, if the photometric period can be confirmed, this would support a nearby low-luminosity object, which would make a UCXB (which should have a high mass transfer rate and therefore high X-ray luminosity) unlikely.

As mentioned in $\S$ 1 slow X-ray pulsars are typically identified with Be or supergiant HMXBs. Pulsars in LMXB systems typically spin much faster (P$_{spin}$ $<$ 200 s) due to the higher spin-up torque provided by Roche lobe overflow, though recently slow pulsars with  P$_{spin}$ $>$ 100 s have been found in symbiotic LMXBs. Therefore if the 626 s period is indicative of the spin of a neutron star, a HMXB is the most likely scenario. Based on simultaneous spectral fitting of \xmm\ EPIC and \emph{INTEGRAL} IBIS data, \citet{mal10} concluded that \puls\ was most likely a HMXB. However, the UKIDS NIR data has a limiting magnitude of $K \sim$ 18 mag, meaning that we are likely to be able to identify all giant type stars of at least Solar mass all the way to the Galactic centre, and we would certainly expect to detect any high mass main sequence stars. For a Galactic centre distance of 8 kpc, the distance modulus is 14.5 mag. We measure an A$_k$ of 1.405 to the Bulge giants in our sample, so any star in that population would be detected with an intrinsic magnitude of M$_k$ $\leq$ 2.1 mag. This equates to all main sequence stars down to and including A type stars, all giants, and all supergiant stars. None of the possible counterparts in the NIR images have photometry or colours consistent with a high mass donor.  A HMXB located on the far side of the Galactic centre (where the NIR counterpart may not be detected due to extinction) is also unlikely, as the neutral hydrogen column density derived from fitting the EPIC spectra is approximately half the Galactic value in that direction \citep[weighted average $n_H$ = 9.69 $\times$ 10$^{21}$ atom cm$^{-2}$;][]{kal05}, and is therefore inconsistent with high levels of extinction. The location of \puls\ on the near side of the Galactic centre is thus more probable, therefore making a HMXB nature unlikely for \puls.

Two of the possible NIR counterparts to \puls\ (sources 1 and 3 in Figure \ref{f:finder}) have colours and photometry consistent with late-type giant stars at distances that put them close to the Galactic centre, although nearby main sequence dwarfs cannot be ruled out with the data at hand. However, the derived spectral types of these candidate counterparts along with the 626 s period and hard X-ray spectrum of \puls\ are all consistent with a symbiotic LMXB containing a slow pulsar located on the near side of the Galactic bulge. Taking a distance of 8.5 kpc, the unabsorbed 0.2 -- 10 keV luminosity of \puls\ (as derived from the best-fit spectral model to the EPIC data) is $\sim$3 $\times$ 10$^{34}$ erg s$^{-1}$, entirely consistent with the known sample of symbiotic X-ray binaries \citep{mas07,nuc07}. It is therefore possible that \puls\ represents a new addition to the select class of symbiotic LMXBs.

\section{Conclusions}

In this paper we have presented analyses of the \xmm\ EPIC data of the source \puls, reporting the discovery of a significant coherent periodic modulation at 626 $\pm$ 2 s with a pulsed fraction in the 0.2 -- 10 keV range of 54 $\pm$ 2\%. Analysis of the EPIC spectra found that the spectrum was most consistent with an absorbed power law model. Phase resolved spectral analyses did not detect any significant variability in the shape of the continuum spectrum over the 626 s modulation. A search through the UKIDSS-GPS DR3 database found four NIR counterparts within the \xmm\ error circle. The extinction corrected photometry of the candidate counterparts and their positions on the colour-magnitude diagrams are consistent with either a local main sequence dwarf (at a distance of $\sim$0.5 -- 1.3 kpc) or a late-type giant in the Galactic bulge.

The 626 s period is consistent with either an orbital modulation in an UCXB or the spin of a compact object. If the compact object is a white dwarf, \puls\ would have to be an AM CVn or an IP located at a distance of $<$ 2 kpc, depending on whether the modulation is orbital or spin in nature. Based on the X-ray spectrum we conclude that \puls\ is unlikely to be an AM CVn. We cannot rule out an IP with the data at hand, although the absence of absorption dips over the 626 s period, a soft BB spectral component, and Fe K$\alpha$ emission lines are abnormal for an IP. 

The 626 s modulation could instead indicate the orbital period of an UCXB, in which case \puls\ would be the most compact UCXB currently known. However, the profile of the folded light curve and the apparent lack of variable absorption with phase are inconsistent with an eclipsing system, therefore likely indicating the presence of a third body which induces an eccentricity in the inner binary orbit. While we cannot rule out that \puls\ is an UCXB, a more likely scenario is that the 626 s period indicates the spin of a neutron star. The lack of a high mass optical/NIR counterpart argues against a HMXB pulsar, leading us to suggest that if the X-ray source is associated with a late-type giant donor star, \puls\ may be a new addition to the rare class of symbiotic LMXBs. Further high spatial resolution X-ray observations and follow-up NIR spectroscopy are required in order to confirm the nature of this system.

\begin{acknowledgements}

We thank the anonymous referee, Phil Evans, Julian Osborne, and Andrew Norton for their comments and discussions which have improved this paper. Based on observations from \xmm,
an ESA science mission with instruments and contributions directly funded by
ESA Member States and NASA. SAF and AJG acknowledge STFC funding. This work made use of the 2XMM Serendipitous
Source Catalogue, constructed by the \xmm\ Survey Science Centre on
behalf of ESA.

\end{acknowledgements}


\begin{thebibliography}{widestentry}


\bibitem[Chakrabarty \& Roche(1997)]{cha97} Chakrabarty, D., \& Roche, P. 1997, ApJ, 489, 254

\bibitem[Chou \& Grindlay(2001)]{ch01} Chou, Y., \& Grindlay, J. E. 2001, ApJ, 563, 934

\bibitem[Corbet et al.(2008)]{cor08} Corbet, R. H. D., Sokoloski, J. L., Mukai, K., Markwardt, C. B., \& Tueller, J. 2008, ApJ, 675, 1424

\bibitem[Cutri et al.(2003)]{cut03} Cutri, R.~M., et al.\ 2003, The IRSA 2MASS All-Sky Point Source Catalog, NASA/IPAC Infrared Science Archive.~http://irsa.ipac.caltech.edu/applications/Gator/, 

\bibitem[Davidsen et al.(1977)]{dav77} Davidsen, A., Malina, R., \& Bowyer, S. 1977, ApJ, 211, 866

\bibitem[Dutra et al.(2003)]{dutr03} Dutra, C.~M., Santiago, B.~X.,
  Bica, E.~L.~D., \& Barbuy, B. 2003, \mnras, 338, 253

\bibitem[Evans \& Hellier(2007)]{eva07} Evans, P. A., \& Hellier, C. 2007, ApJ, 663, 1277

\bibitem[Garcia et al.(1983)]{gar83} Garcia, M. R., Baliunas, S. L., Doxsey, R., et al. 1983, ApJ, 267, 291

\bibitem[Gosling et al.(2007)]{gosl07} Gosling, A.~J., Bandyopadhyay,
  R.~M., Miller-Jones, J.~C.~A., \& Farrell, S.~A. 2007, \mnras, 380,
  1511

\bibitem[Gosling et al.(2009)]{gosl09} Gosling, A.~J., Bandyopadhyay,
  R.~M., \& Blundell, K.~M.\ 2009, \mnras, 394, 2247

\bibitem[Halpern \& Gotthelf(2010a)]{hal10a} Halpern, J. P., \& Gotthelf, E. V. 2010a, ATEL, 2664, 1

\bibitem[Halpern \& Gotthelf(2010b)]{hal10b} Halpern, J. P., \& Gotthelf, E. V. 2010b, ATEL, 2681, 1

\bibitem[Hellier \& Mukai(2004)]{hel04} Hellier, C., \& Mukai, K.\ 2004, \mnras, 352, 1037 


\bibitem[Ikhsanov(2007)]{ikh07} Ikhsanov, N. R., 2007, MNRAS, 375, 698

\bibitem[in 't Zand et al.(2007)]{int07} in 't Zand, J. J. M., Jonker, P. G., \& Markwardt, C. B. 2007, A\&A, 465, 953

\bibitem[Jansen et al.(2001)]{jan01} Jansen, F., Lumb, D., Altieri, B., et al. 2001, A\&A, 365, L1

\bibitem[Kalberla et al.(2005)]{kal05} Kalberla, P. M. W., Burton, W. B., Hartmann, D. et al. 2005, A\&A, 440, 775

\bibitem[Kaplan et al.(2007)]{kap07} Kaplan, D. L., Levine, A. M., Chakrabarty, D., et al. 2007, ApJ, 661, 437

\bibitem[Kong et al.(1998)]{ko98} Kong, A.~K.~H., Charles,
P.~A., \& Kuulkers, E. 1998, \na, 3, 301

\bibitem[Koyama et al.(1991)]{koy91} Koyama, K., Kunieda, H., Takeuchi, Y., \& Tawara, Y. 1991, ApJ, 370, L77

\bibitem[Kuiper et al.(2002)]{kui02} Kuiper, L., Hermsen, W., Verbunt, F., Ord, S., Stairs, I., \& Lyne, A. 2002, ApJ, 577, 917

\bibitem[Kuulkers et al.(2006)]{ku06} Kuulkers, E., Norton, A., Schwope, A., \& Warner, B. 2006, X-rays from Cataclysmic Variables, in Compact Stellar X-ray Sources, eds. W. Lewin \& M. van der Klis (Cambridge Univ. Press, Cambridge), 421

\bibitem[Larsson(1996)]{lar96} Larsson, S.\ 1996, \aaps, 117, 197 

\bibitem[Lawrence et al.(2007)]{lawr07} Lawrence, A., Warren, S. J., Almaini, O., et al. 2007,
  \mnras, 379, 1599

\bibitem[Lewin et al.(1971)]{lew71} Lewin, W. H. G., Ricker, G. R., \& McClintock, J. E. 1971, ApJ, 169, L17

\bibitem[Li et al.(2008)]{li08} Li, X.-H., Lu, F.-J, \& Li, Z. 2008, ApJ, 682, 1166

\bibitem[Liu et al.(2001)]{liu01} Liu, Q. Z., van Paradijs, J., \& van den Heuvel, E. P. J. 2001, A\&A, 368, 1021

\bibitem[Liu et al.(2006)]{liu06} Liu, Q. Z., van Paradijs, J., \& van den Heuvel, E. P. J. 2006, A\&A, 455, 1165

\bibitem[Lomb(1975)]{lo75} Lomb, N. R. 1975, Ap$\&$SS, 39, 447

\bibitem[Lucas et al.(2007)]{luca07} Lucas, P.~W., Hoare, M. G., Longmore, A., et al. 2008, MNRAS, 391, 136

\bibitem[Monet et al.(2003)]{mon03} Monet, D. G., et al. 2003, AJ, 125, 984

\bibitem[Malizia et al.(2010)]{mal10} Malizia, A., Bassani, L, Sguera, V., Stephen, J. B., Bazzano, A., Fiocchi, M., Bird, A. J., 2010, MNRAS, accepted, arXiv:1006.1272

\bibitem[Masetti et al.(2002)]{mas02} Masetti, N., Dal Fiume, D., Cusumano, G., et al. 2002, A\&A, 382, 104

\bibitem[Masetti et al.(2006a)]{mas06a} Masetti, N., Orlandini, M., Palazzi, E., \& Amati, L., Frontera, F. 2006a, A\&A, 453, 295

\bibitem[Masetti et al.(2006b)]{mas06b} Masetti, N., Rigon, E., Maiorano, E., et al. 2006b, A\&A, 464, 277

\bibitem[Masetti et al.(2007)]{mas07} Masetti, N., Landi, R., Pretorius, M. L., et al. 2007, A\&A, 470, 331

\bibitem[Mattana et al.(2006)]{mat06} Mattana, F., G\"{o}tz, D., Falanga, M., et al. 2006, A\&A, 460, L1

\bibitem[Nagase(1989)]{nag89} Nagase, F. 1989, PASJ, 41, 1

\bibitem[Nucita et al.(2007)]{nuc07} Nucita, A. A., Carpano, S., \& Guainazzi, M. 2007, A\&A, 474, L1

\bibitem[Press \& Rybicki(1989)]{pr89} Press, W. H., \& Rybicki, G. B. 1989, ApJ, 338, 277

\bibitem[Ramsay et al.(2005)]{ra05} Ramsay, G., Hakala, P., Marsh, T., Nelemans, G., Steeghs, et al. 2005, A\&A, 440, 675

\bibitem[Rieke \& Lebofsky(1985)]{riek85} Rieke, G.~H., \& Lebofsky,
  M.~J. 1985, \apj, 288, 618

\bibitem[Ritter \& Kolb(2003)]{rit09} Ritter, H., \& Kolb, U. 2003, A\&A, 404, 301
  
\bibitem[Sakano et al.(2002)]{sak02} Sakano, M., Koyama, K., Murakami, H., Maeda, Y., \& Yamauchi, S. 2002, ApJS, 138, 19

\bibitem[Scargle(1982)]{sc82} Scargle, J. D.
1982, ApJ, 263, 835

\bibitem[Sidoli et al.(2001)]{sid01} Sidoli, L., Belloni, T., \& Mereghetti, S. 2001, A\&A, 368, 835




\bibitem[Voges et al.(1999)]{vog99} Voges, W., Aschenbach, B., Boller, Th., et al. 1999, \aap,
349, 389

\bibitem[Voges et al.(2000)]{vog00} Voges, W., Aschenbach, B., Boller, Th., et al. 2000,
\iaucirc, 7432, 1

\bibitem[Warner(1995)]{war95} Warner, B. 1995, Cataclysmic Variable Stars (Cambridge Univ. Press, Cambridge)

\bibitem[Watson et al.(2009)]{wat08} Watson, M.~G., Schr\"{o}der, A. C., Fyfe, D. et al.
2009, A\&A, 493, 339


\bibitem[White \& Priedhorsky(1986)]{whi86} White, N.~E., \& Priedhorsky, W.\ 1986, \baas, 18, 1048 


\bibitem[Zdziarski et al.(2007)]{zdz07} Zdziarski, A. A., Wen, L., \& Gierlinski, M. 2007, MNRAS, 377, 1006

\end{thebibliography}
\end{document}